\documentclass[prA,floatfix,twocolumn,amsmath]{revtex4}
\usepackage{amsthm}
\usepackage{bbm}
\usepackage{amssymb}
\usepackage{verbatim}
\usepackage{graphicx}
\usepackage{epsfig}

\usepackage{graphicx}
\usepackage{pb-diagram}
\usepackage{lamsarrow}
\usepackage{pb-lams}

\newtheorem{thm}{Theorem}
\newtheorem*{thm*}{Theorem}
\newtheorem{prop}[thm]{Proposition}

\newtheorem{lem}[thm]{Lemma}
\theoremstyle{definition}

\theoremstyle{remark}

\newcommand{\lra}{\longrightarrow}

\newcommand{\be}{\begin{equation}}
\newcommand{\ee}{\end{equation}}
\newcommand{\bea}{\begin{eqnarray}}
\newcommand{\eea}{\end{eqnarray}}

\newcommand{\1}{\mathbbm{1}}
\newcommand{\tr}[1]{{\rm tr}\left[#1\right]}

\DeclareMathOperator{\range}{range}
\DeclareMathOperator{\gap}{gap}
\newcommand{\C}{{\cal C}}
\renewcommand{\>}{\rangle}
\newcommand{\<}{\langle}

\begin{document}

\title{PEPS as unique ground states of local Hamiltonians}

\author{D. P\'{e}rez-Garc\'{\i}a}
\affiliation{Departamento de Analisis Matematico. Universidad
Complutense de Madrid, 28040 Madrid, Spain.}

\author{F. Verstraete}
\affiliation{Fakult\"{a}t f\"{u}r Physik, Universit\"{a}t Wien, Boltzmanngasse
5, A-1090 Wien, Austria.}

\author{J.I. Cirac}
\affiliation{Max Planck Institut f\"{u}r Quantenoptik,
Hans-Kopfermann-Str. 1, Garching, D-85748, Germany.}

\author{M.M. Wolf}
\affiliation{Max Planck Institut f\"{u}r Quantenoptik,
Hans-Kopfermann-Str. 1, Garching, D-85748, Germany.}

\begin{abstract}
In this paper we consider projected entangled pair states (PEPS)
on arbitrary lattices. We construct local parent Hamiltonians for
each PEPS and isolate a condition under which the state is the
unique ground state of the Hamiltonian. This condition, verified
by generic PEPS and examples like the AKLT model, is an injective
relation between the boundary and the bulk of any local region.
While it implies the existence of an energy gap in the 1D case we
will show that in certain cases  (e.g., on a 2D hexagonal lattice)
the parent Hamiltonian can be gapless with a critical ground
state. To show this we invoke a mapping between classical and
quantum models and prove that in these cases the injectivity
relation between boundary and bulk solely depends on the lattice
geometry.
\end{abstract}


\maketitle

\section{Introduction}

Exactly solvable models for quantum mechanical spin systems are
important testbeds and sources of insight for the theory of
condensed matter systems and their statistics. Unfortunately, such
models are rare even in 1D and, disregarding free particles,
essentially non-existent in higher spatial dimensions. If the
focus lies in low temperature physics then quasi-exactly solvable
models for which at least the ground state is
 known exactly become interesting. A powerful tool for constructing such models in 1D
is the matrix product state (MPS) formalism
\cite{MPS-Perez,FaNaWe,MPS} which has his roots in the AKLT model
\cite{AKLT} and has experienced a remarkable development in the
last years concerning both analytical results \cite{ref-MPSa} and
numerical methods \cite{ref-MPSn}. In two and higher spatial
dimensions a generalization of the MPS idea is known as projected
entangled pair state (PEPS) formalism
\cite{PEPS-Frank-0,PEPS-Frank,PEPS-Norbert}, earlier attempts of
which can already be found in the seminal work of AKLT \cite{AKLT}
as well as in studies of dimer models \cite{dimer} and
antiferromagnetic vertex spin models \cite{vertex}. The idea of
the PEPS representation of quantum states has its roots in quantum
information theory \cite{cluster-Frank} but it found most of its
applications in quantum many-body physics
\cite{PEPS-Frank-0,PEPS-Frank,PEPS-Norbert,Valentin,Hastings}.

The present work will continue on this line and investigate
quasi-exactly solvable models based on PEPS. Our main interest
lies hereby in the uniqueness of the ground state. We prove that
\begin{itemize}
    \item every PEPS satisfying a condition we call 'injectivity'
    is the unique ground state of a local Hamiltonian.
    \item For quantum states which are coherent versions of
    classical Gibbs states we show that injectivity depends solely
    on the lattice geometry.
    \item By invoking the mentioned classical-to-quantum mapping \cite{PEPS-Frank} we
    construct a PEPS which is the unique critical ground state of
    a local Hamiltonian on a 2D hexagonal lattice. This shows
    that, in contrast to 1D MPS, injectivity does not imply the
    existence of an energy gap.
    \item We give an example on the 2D square lattice where
    uniqueness can be proven in the absence of injectivity.
    \item We provide a computable sufficient condition for the
    existence of an energy gap.
    \item In the appendix we show that every translational
    invariant PEPS on a 2D square lattice admits a representation
    with site-independent tensors.
\end{itemize}

Note that the relation between injectivity and uniqueness
generalizes in a fairly direct way what is known in the 1D case
\cite{MPS-Perez,FaNaWe} to arbitrary lattice geometries. The main
difference lies in the fact that beyond the 1D case the existence
of a gap above the ground state energy becomes a rather
independent property, the determination of which essentially
remains an open problem.


\section{Preliminaries}\label{sec:prel}

Let us start by recalling that a PEPS is a quantum state
$|\varphi\>$ constructed over a graph (Fig.1) such that each
vertex corresponds to a physical site and the geometry of the
graph typically
 resembles the spatial geometry of a lattice. To construct the PEPS we assign to each edge (bond) of
the graph a maximally entangled state $\sum_{i=1}^D|ii\>$. In this
way, each vertex gets assigned to as many $D$-dimensional systems
as there are adjacent edges---we may think of these as a
\emph{virtual} substructure. To obtain the final state we apply a
map $P:\C^D\otimes\cdots\otimes \C^D\lra \C^d$ at each vertex
which maps the virtual onto the physical system. Note that $P$ and
$d$ may depend on the vertex and $D$ on the edge. For the sake of
simplicity we will, however, choose the dimensions $D$ and $d$ to
be constant over the entire graph. Let us denote by $e_v$ the
number of edges at vertex $v$ and parameterize the corresponding
map $P^{(v)}$ by
$$P^{(v)}=\sum_{i=1}^d\sum_{j_1,\ldots,j_{e_v}=1}^D A^{(v)}_{i,j} |i\rangle\langle
j_1,\ldots,j_{e_v}|$$such that each $A^{(v)}_i$ is a tensor with
$e_v$ indices $j=(j_1,\ldots,j_{e_v})$. From this we obtain a PEPS
of the form \be |\varphi\rangle=\sum_{i_1,i_2,\ldots=1}^d{\cal
C}\left[\{A^{(v)}_{i_v}\}_v\right]|i_1,i_2,\ldots\rangle,\ee where
$\cal C$ means the contraction of all tensors $A^{(v)}_i$
according to the edges of the graph. If, for instance, we deal
with an $N\times M$ square lattice on a torus we may specify each
vertex by the respective row $j$ and column $k$ such that
\begin{equation}\label{eq:PEPS-noTI}
|\varphi\rangle=\sum_{i_{(1,1)},\ldots,i_{(N,M)}=1}^d {\cal
C}\left[\{A^{(j,k)}_{i_{(j,k)}}\}_{(j,k)}\right]|i_{(1,1)}\cdots
i_{(N,M)}\rangle
\end{equation}
where each $A^{(j,k)}_i$ is now a $4$-index tensor which is
contracted according to the lattice:
$$
\begin{diagram}\dgARROWLENGTH=1em
\node{}  \node{} \arrow{s,-} \node{} \arrow{s,-}
\node{}\\
\node{} \arrow{l,-}\node{\fbox{$A^{(1,1)}_{i_{(1,1)}}$}}
\arrow{s,-}\arrow{l,-} \node{\fbox{$A^{(1,2)}_{i_{(1,2)}}$}}
\arrow{s,-}\arrow{l,-}
\node{\cdots}\\
\node{} \arrow{l,-} \node{\fbox{$A^{(2,1)}_{i_{(2,1)}}$}}
\arrow{s,-}\arrow{l,-} \node{\fbox{$A^{(2,2)}_{i_{(2,2)}}$}}
\arrow{s,-}\arrow{l,-}
\node{\cdots}\\
\node{}\node{\cdots} \node{\cdots}
\end{diagram}
$$

Clearly, if the tensors $A^{(j,k)}_i$ do not depend on the vertex
$v=(j,k)$ then the corresponding PEPS $|\varphi\rangle$ is
translational invariant. Conversely, if $|\varphi\rangle$ is
translational invariant, then there  exists always a
representation (with presumably larger $D$) such that
$A^{(j,k)}_i=A_i$ is site-independent. A proof of this statement
can be found in the appendix. We will throughout not assume such a
symmetry and allow the $A_i$'s to be site dependent though we
typically omit to write the explicit dependence $A^{(v)}_i$.

\section{Parent Hamiltonians}\label{sec:ham}

In the following we will mainly consider lattices instead of
arbitrary graphs as this gives us a meaningful notion of locality
and ensures that  volumes grow faster than boundary areas. The
latter is, in fact, the main property which allows us to construct
parent Hamiltonians for every PEPS $|\varphi\rangle$. Consider a
region $R$ on the lattice with respective reduced density operator
$\rho_R={\rm tr}_{\backslash R}|\varphi\rangle\langle\varphi|$.
The support space $S_R$ of $\rho_R$ has dimension at most
$D^{e_R}$ where $e_R$ is the number of edges connecting the
interior of $R$ to the exterior. Denote by $|R|$ the number of
lattice sites in $R$ with total Hilbert space ${\cal
H}_R=\mathbb{C}^{d\otimes|R|}$. Since $S_R$ grows as the boundary
of $R$ whereas ${\cal H}_R$ grows like its volume, we can always
find a sufficiently large region such that $\rho_R$ will have a
non-trivial kernel ${\rm ker}(\rho_R)={\cal H}_R\setminus S_R$. It
is thus a ground state of every Hamiltonian $h$ with the following
properties
\begin{itemize}
\item $h\ge 0$, \item $\ker(h)=S_R$,
\end{itemize}
since by construction $\tr{\rho_R h}=0$. This Hamiltonian can then
be extended to the whole lattice by assigning such a surrounding
region $R_v$ and a corresponding $h_v$ to every vertex $v$ and
defining \be \label{eq:H1} H=\sum_{v} h_v\otimes\1_{\backslash
R_v}.\ee In this way we have again $H\geq 0$ and
$\langle\varphi|H|\varphi\rangle=\sum_v\tr{\rho_{R_v}h_v}=0$. The
Hamiltonian is thus {\it local} and {\it frustration free}
\cite{MPS-Perez,FaNaWe}, in the sense that $|\varphi\>$ minimizes
the energy locally for every interaction term $h_v$. The state is
therefore sometimes called an {\it optimum ground state}
\cite{vertex}. We remark that it was recently proven in
\cite{Hastings} that all local gapped Hamiltonians can be
efficiently approximated by frustration free ones.

\begin{figure}[tttt]
\begin{center}
\epsfig{file=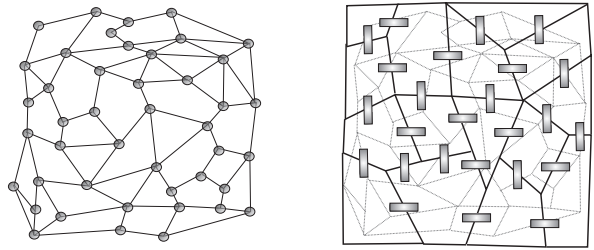,angle=0,width=0.9\linewidth}
\end{center}
\caption {Left: A PEPS can be assigned to an arbitrary graph. Each
vertex corresponds to a physical site and every edge depicts the
contraction of tensors assigned to the vertices. Right: A parent
Hamiltonian which has the PEPS as unique ground state can be
constructed by first joining vertices in injective regions (now
surrounded by solid lines) and defining interactions (depicted by
bars) between connected neighboring regions. } \label{figure}
\end{figure}

\section{Injectivity}\label{sec:inj}

We will now discuss the main condition which will later on turn
out to be sufficient for proving uniqueness of the ground state
for  Hamiltonians of the form in Eq.(\ref{eq:H1}). Consider a PEPS
$|\varphi\>$ given by tensors $A_i$ (depending on the vertex). Let
$R$ be a connected region on the graph containing $|R|$ sites
(vertices) and having $e_R$ outgoing edges (bonds). The linear map
$\Gamma_R:\mathbb{C}^{D^{e_R}}\rightarrow\mathbb{C}^{d^{|R|}}$
defined by
\be\Gamma_R(C)=\sum_{i_R}\C\left[A_{i_R}C\right]|i_R\>\ee maps the
virtual boundary of $R$ to the physical bulk. Here $A_{i_R}$
stands for the tensor obtained by contracting all the $A_i$'s in
the region $R$ with  bonds open in the boundary of $R$ (and then
$C$ is a tensor with the same number $e_R$ of bonds). Graphically
we will represent $\C[A_{i_R}C]$ as
$$
\begin{diagram}\dgARROWLENGTH=1em
\node{\fbox{$A_{i_R}$}}\arrow{l,-} \node{\fbox{$C$}}
\end{diagram}
$$
where we have joined all the $e_R$ open bonds of $A_{i_R}$ and $C$
in one line.

If $\Gamma_R$ is an injective map, we will simply say that the
region $R$ is injective. Similarly, if there is a proper covering
of disjoint injective regions for the entire graph we will call
the PEPS itself injective.

Of course, the injectivity property of $R$ is equivalent to the
fact that the finite set $\{A_{i_R}\}_{i_R}$ generates the space
of tensors with $e_R$  $D$-dimensional indices. Just counting
degrees of freedom one can see that the dimension of the space of
tensors that we have to generate grows like the boundary
$D^{e_R}$, whereas the number of tensors that we have grows like
the volume $d^{|R|}$. This strongly suggest that the injectivity
condition is {\it generic} on any lattice, i.e., almost always
fulfilled as long as $d^{|R|}\geq D^{e_R}$. Indeed, if one takes a
numerically random  PEPS, it appears to be fulfilled in all the
cases. As a more concrete example, one can also check that the 2D
AKLT state \cite{AKLT} is injective.

In the remaining part of this section we will prove a pair of
Lemmas which are particularly concerned with lifting the
injectivity property and its consequences from smaller to larger
regions. The reader not interested in the technical details may
skip this part.

\begin{lem}\label{lem1}
If $R$ and $S$ are two disjoint injective regions, then $R\cup S$
is also injective.
\end{lem}

\begin{proof}

We start from the picture
$$
\begin{diagram}\dgARROWLENGTH=1em
\node{\fbox{$A_{i_R}$}}\arrow{e,t,-}{a} \arrow{s,l,-}{b} \\
\node{\fbox{$A_{i_S}$}}\arrow{e,t,-}{c}
\end{diagram}
$$

Since both $A_{i_R}$ and $A_{i_S}$ generate the respective space
of tensors, then, summing in $i_R$ and $i_S$ we can get
$$
\begin{diagram}\dgARROWLENGTH=1em
\node{\fbox{$\delta_{a,a_0}$}}\arrow{e,t,-}{a} \arrow{s,l,-}{b}
\\
 \node{\fbox{$\delta_{c,c_0}$}}\arrow{e,t,-}{c}
\end{diagram}
$$
for each $a_0$ and $c_0$. Hence, we have injectivity in the region
$R\cup S$, since we can generate all the canonical basis vectors
of the respective space of tensors.
\end{proof}

Let us call $G_R=\range(\Gamma_R)$ the subspace generated in the
physical bulk by acting on the virtual boundary. Note that the
 reduced density operator $\rho_R$ is supported in $G_R\supseteq
 S_R$. Moreover, if $R$ and its complement are injective then the
 converse holds as well so that $G_R=S_R$.

\begin{lem}\label{lem:crucial}
Let $R_1, R_2, R_3$ be three disjoint regions, and
$\mathcal{H}_R=\mathbb{C}^{d^{|R|}}$ denote the total Hilbert
space corresponding to region $R$.
\begin{enumerate}
    \item $G_{R_1\cup R_2\cup R_3} \subseteq \left(G_{R_1\cup R_2}\otimes \mathcal{H}_{R_3}\right) \cap \left(\mathcal{H}_{R_1}\otimes G_{R_2\cup
    R_3}\right)$.
    \item If $R_1$ and $R_3$ are not connected and $R_2$ and $R_3$
    are both injective, then $G_{R_1\cup R_2\cup R_3} = \left(G_{R_1\cup R_2}\otimes \mathcal{H}_{R_3}\right) \cap \left(\mathcal{H}_{R_1}\otimes G_{R_2\cup
    R_3}\right)$.
    \item If all three regions are injective, then $G_{R_1\cup R_2\cup R_3} = \left(G_{R_1\cup R_2}\otimes \mathcal{H}_{R_3}\right) \cap \left(\mathcal{H}_{R_1}\otimes G_{R_2\cup
    R_3}\right) \cap \left(G_{R_1\cup R_3}\otimes
    \mathcal{H}_{R_2}\right)$.
\end{enumerate}
\end{lem}

\begin{proof}
It is clear that $|\psi\>\in G_{R_1\cup R_2}\otimes
\mathcal{H}_{R_3}$ if and only if $|\psi\>$ can be written as
$\sum_{i_{R_3}}\Gamma_{R_1\cup R_2}(C_{i_{R_3}})|i_{R_3}\>$. Then,
if $|\psi\>\in G_{R_1\cup R_2\cup R_3}$, since  we can write
$|\psi\>=\sum_{i_{R_1}, i_{R_2}, i_{R_3}}\C(A_{R_1} A_{R_2}
A_{R_3} X)|i_{R_1}i_{R_2}i_{R_3}\>$, calling
$C_{i_{R_3}}=\C(A_{i_{R_3}}X)$, we have that $|\psi\>\in
G_{R_1\cup R_2}\otimes \mathcal{H}_{R_3}$.

Of course, the same can be done for $\mathcal{H}_{R_1}\otimes
G_{R_2\cup R_3}$, which proves part 1. of the Lemma.

For the other parts note that if $|\psi\>\in\left(G_{R_1\cup
R_2}\otimes \mathcal{H}_{R_3}\right) \cap
\left(\mathcal{H}_{R_1}\otimes G_{R_2\cup R_3}\right)$, then
$$|\psi\>=\sum_{i_{R_3}}\Gamma_{R_1\cup R_2}(D_{i_{R_3}})|i_{R_3}\>=
\sum_{i_{R_1}}\Gamma_{R_2\cup R_3}(C_{i_{R_1}})|i_{R_1}\>,$$ which
implies that
$$\begin{diagram}\dgARROWLENGTH=1em
\node{}\arrow{s,-} \arrow[2]{e,t,-}{u} \node[2]{} \arrow[2]{s,-} \\
\node{\fbox{$A_{i_{R_2}}$}}\arrow[2]{s,l,-}{d} \arrow{see,t,-}{l} \\
\node{} \node[2]{\fbox{$C_{i_{R_1}}$}}\\
\node{\fbox{$A_{i_{R_3}}$}} \arrow{nee,t,-}{x} \arrow{s,-}\\
\node{}\arrow[2]{e,b,-}{y}\node[2]{} \arrow[2]{n,-}
\end{diagram}
\quad = \quad
\begin{diagram}\dgARROWLENGTH=1em
\node{}\arrow{s,-} \arrow[2]{e,t,-}{a} \node[2]{} \arrow[2]{s,-} \\
\node{\fbox{$A_{i_{R_1}}$}}\arrow[2]{s,l,-}{u} \arrow{see,t,-}{b} \\
\node{} \node[2]{\fbox{$D_{i_{R_3}}$}}\\
\node{\fbox{$A_{i_{R_2}}$}} \arrow{nee,t,-}{l} \arrow{s,-}\\
\node{}\arrow[2]{e,b,-}{d}\node[2]{} \arrow[2]{n,-}
\end{diagram}
$$
for every $i_{R_1}, i_{R_2}, i_{R_3}$. Or equivalently (summing in
the repeated indices),
\begin{equation}\label{eq.1}
A_{i_{R_2}}^{u,l,d}A_{i_{R_3}}^{d,x,y}C_{i_{R_1}}^{u,l,x,y}=
A_{i_{R_1}}^{a,b,u}A_{i_{R_2}}^{u,l,d}D_{i_{R_3}}^{a,b,l,d},
\end{equation} where the sum over $a$ and $y$ is only non-trivial if $R_1$ is not separated from $R_3$.

In order to prove 2. we now use injectivity in $R_2$ and $R_3$.
That is, for every fixed set of indices $l_0,u_0,d_0$ and
$x_0,y_0$ there exist $\alpha_{i_{R_2}}^{u_0,l_0,d_0}$ and
$\omega_{i_{R_3}}^{x_0,y_0}$ such that
\begin{eqnarray}\label{eq.2}
\sum_{i_{R_2}} \alpha_{i_{R_2}}^{u_0,l_0,d_0} A_{i_{R_2}}^{u,l,d}
&=&\delta_{u,u_0}\delta_{l,l_0}\delta_{d,d_0},\\
\sum_{i_{R_3}} \omega_{i_{R_3}}^{x_0,y_0} A_{i_{R_3}}^{d,x,y}
&=&\delta_{x,x_0}\delta_{y,y_0}.
\end{eqnarray}
Inserting this into Eq.(\ref{eq.1}) we get that
\begin{equation*}
C_{i_{R_1}}^{u,l,x,y}=
A_{i_{R_1}}^{a,b,u}\underset{M^{a,b,l,x,y}}{\underbrace{\sum_{i_{R_3}}\omega_{i_{R_3}}^{x,y}D_{i_{R_3}}^{a,b,l,d}}},
\end{equation*}
(note that $M^{a,b,l,x,y}$ does not depend on $d$). Hence, the
coefficients of $\psi$ have the form of the left part of the
following identity (proven below)
\begin{equation}
\begin{diagram}\dgARROWLENGTH=1em
\node{}\arrow{s,-} \arrow[2]{e,t,-}{a} \node[2]{} \arrow[2]{s,-} \\
\node{\fbox{$A_{i_{R_1}}$}} \arrow{see,t,-}{b} \arrow{s,l,-}{u}\\
\node{\fbox{$A_{i_{R_2}}$}}\arrow{s,l,-}{d} \arrow[2]{e,t,-}{l} \node[2]{\fbox{$M$}}\\
\node{\fbox{$A_{i_{R_3}}$}} \arrow{nee,t,-}{x} \arrow{s,-}\\
\node{}\arrow[2]{e,b,-}{y}\node[2]{} \arrow[2]{n,-}
\end{diagram}\quad = \quad \begin{diagram}\dgARROWLENGTH=1em
\node{}\arrow{s,-} \arrow[2]{e,t,-}{w} \node[2]{} \arrow[2]{s,-} \\
\node{\fbox{$A_{i_{R_2}}$}} \arrow{see,t,-}{l} \arrow{s,l,-}{d}\\
\node{\fbox{$A_{i_{R_3}}$}}\arrow{s,l,-}{z} \arrow[2]{e,t,-}{x} \node[2]{\fbox{$\tilde{M}$}}\\
\node{\fbox{$A_{i_{R_1}}$}} \arrow{nee,t,-}{b} \arrow{s,-}\\
\node{}\arrow[2]{e,b,-}{v}\node[2]{} \arrow[2]{n,-}
\end{diagram}\label{eq:identity1}
\end{equation}
which implies that under the conditions of part 2. $|\psi\>\in
G_{R_1\cup R_2 \cup R_3}$ completing the proof of the second part.

For 3. we first repeat the above proof with a cyclic permutation
of the regions, i.e., $(R_1,R_2,R_3)\rightarrow (R_2,R_3,R_1)$
leading to the identity in (\ref{eq:identity1}). Then we use
injectivity in all three regions in order to replace
$A_{i_{R_1}}^{a,b,u}$, $A_{i_{R_2}}^{u,l,d}$ and
$A_{i_{R_3}}^{d,x,y}$ by $\delta_{a,a_0}\delta_{b,b_0}$,
$\delta_{l,l_0}$ and $\delta_{x,x_0}\delta_{y,y_0}$ respectively.
Inserting this into Eq.(\ref{eq:identity1}) we obtain $
M^{a,b,l,x,y}=\delta_{a,y} \tilde{M}^{w,l,x,b,v}$ so that
$$
|\psi\>=\sum_{i_{R_1},i_{R_2},i_{R_3}}\quad \quad
\begin{diagram}\dgARROWLENGTH=1em
\arrow[4]{s,-}\arrow{e,-}\\
\node{\fbox{$A_{i_{R_1}}$}} \arrow{n,-}\arrow{s,-}\arrow{see,-} \\
\node{\fbox{$A_{i_{R_2}}$}}\arrow{s,-}\arrow[2]{e,-} \node[2]{\fbox{$\tilde{M}$}}\\
\node{\fbox{$A_{i_{R_3}}$}} \arrow{s,-}\arrow{nee,-}\\
\arrow{e,-}
    \end{diagram}\quad |i_{R_1},i_{R_2},i_{R_3}\>
    $$
is indeed in $G_{R_1\cup R_2 \cup R_3}$.
\end{proof}

\section{Uniqueness of the ground state}\label{sec:par}

Our aim now is to prove that every injective $|\varphi\>$ is the
unique ground state of a Hamiltonian of the form in
Eq.(\ref{eq:H1}). Consider an injective tile of the lattice, i.e.,
a covering by disjoint injective regions and let us merge sites
within each injective region such that in the new, regrouped
lattice each single vertex is injective. We connect the vertices
in this super-lattice if the respective regions were connected
before (see Fig.1). As a Hamiltonian we choose
\be\label{Hab}H=\sum_{(\alpha,\beta)}
h_{\alpha,\beta}\otimes\1,\ee where the sum runs over all edges
$(\alpha,\beta)$ in the super-lattice and $h_{\alpha,\beta}\geq 0$
is any nearest-neighbor interaction term satisfying
$\ker(h_{\alpha,\beta})=G_{R_\alpha\cup R_\beta}$.

\begin{thm}
\label{thm:main} For every injective PEPS on an arbitrary lattice
there is a local frustration-free Hamiltonian such that the state
is its unique ground state.
\end{thm}
{\it Remark:} Note that though  Eq.(\ref{Hab}) gives a standard
construction this Hamiltonian is not unique. We may in particular
choose $h_{\alpha,\beta}'=h_{\alpha,\beta} +
h_{\alpha,\beta}Qh_{\alpha,\beta}$ for any $Q\geq 0$. Similarly,
we can use the same construction for larger regions, or sometimes
even for smaller ones for which injectivity does not yet hold (see
below). The advantage of the latter is to reduce the number of
sites involved in the interaction, as done for instance in the
AKLT model.
\begin{proof}
Using the Hamiltonian of Eq.(\ref{Hab}) and writing, with a slight
abuse of notation, $G_R$ for $G_R\otimes {\cal H}_{\backslash R}$
we have to show that \be\label{eq:2show} \bigcap_{(\alpha,\beta)}
G_{R_\alpha\cup R_\beta}= G_{\cup_\alpha R_\alpha}\ee as the left
hand side is the ground state space of $H$ and the right hand side
is the one-dimensional space corresponding to the PEPS. We will
show this identity by induction. To this end consider a collection
of disjoint regions $\cup_{i=0}^N A_i =A$ such that
$A_1,\ldots,A_N$ are connected to a region $B$ (all regions being
injective). Then as $G_A\subseteq G_{A_0\cup A_i}$ we have that
\be G_A\bigcap_{i=1}^N G_{A_i\cup B} \subseteq
\bigcap_{i=1}^N\Big(G_{A_0\cup A_i}\cap G_{A_i\cup
B}\Big)=\bigcap_{i=1}^N G_{A_0\cup A_i\cup B},\ee where the last
equality follows from part 2. of Lemma \ref{lem:crucial}. If we
exploit in addition that $G_A\subseteq G_{\cup_{i\in I}A_i}$ for
any $I\subseteq\{0,\ldots,N\}$ together with part 3. of Lemma
\ref{lem:crucial} we obtain \be G_A\bigcap_{i=1}^N G_{A_i\cup B}
\subseteq G_{A\cup B}.\label{eq:ind}\ee Eq.(\ref{eq:ind}) can now
be used as an induction step in order to show that the l.h.s. of
Eq.(\ref{eq:2show}) is contained in the r.h.s. . The converse
inclusion follows, however, from part 1. of Lemma
\ref{lem:crucial} which completes the proof.
\end{proof}

\section{Quantum states from classical models}\label{sec:pep}

In \cite{PEPS-Frank} we constructed  PEPS from any classical spin
model with nearest-neighbor interaction (with $d$ possible
configurations per site and a fixed given temperature
$\frac{1}{\beta}$) in such a way that (i) the physical dimension
coincides with the bond dimension, i.e., $d=D$; and (ii) the PEPS
reproduces the correlations of the classical thermal state.  In a
sense, this map replaces thermal by quantum fluctuations. In this
section we will analyze in which cases PEPS constructed in this
way are unique ground states of their parent Hamiltonians from
Eq.(\ref{Hab}).

First of all we will review the construction of \cite{PEPS-Frank}.
For a classical Hamiltonian of the form $H(x)=\sum_{(i,j)}h(x_i,
x_j)$, with $x_i=1,\ldots,d$ we define the associate PEPS as
$$|\psi\>=\frac{1}{\sqrt{Z}}\exp\left[-\frac{\beta}{2}\sum_{(i,j)}\hat{h}_{ij}\right]|+\cdots+\>,$$
where $|+\>=\sum_{x=1}^d|x\>$ and $\hat{h}_{ij}$ is a diagonal
operator that acts as $\hat{h}_{ij}|x_ix_j\>=h(x_i,x_j)|x_ix_j\>$.
To see explicitly the PEPS structure of $|\psi\>$ one can notice
that the non-unitary gate $\exp[-\frac{\beta}{2}\hat{h}_{ij}]$ is
indeed equal to $P_i\otimes P_j |I\>$, where
$|I\>=\sum_{x=1}^d|xx\>$ is an auxiliary maximally entangled state
between the sites $i$ and $j$ and $P_i:\C^{d^2}\lra \C^d$ is an
operator (from the joint physical-virtual system into the physical
system) acting as $P_i|x_iv\>=|x_i\> \<\phi^i_{x_i}|v\>$. Here the
vectors $|\phi^i_{x_i}\>$ can be obtained by a SVD to verify
$\exp[-\frac{\beta}{2}h(x_i,x_j)]=\sum_{k}\<\phi^i_{x_i}|k\>\<k|\phi^j_{x_j}\>$.
Then, the tensors defining the PEPS $|\psi\>$ are
\begin{equation}\label{eq:tensor-classical} A^i_{x_i;\alpha_1\ldots\alpha_n} = \prod_{e}
\<\phi^e_{x_i}|\alpha_e\> \;,\quad \alpha_e=1,\ldots,
d\;,\end{equation} where $|\phi^e_{x_i}\>$ may, of course, depend
on the lattice site $i$, and the product is taken over all edges
$e$ connected with this site. Generically, $\phi^e$ can be
considered (via $\<x_i|\mapsto \<\phi^e_{x_i}|$) an
\emph{invertible} $d\times d$ matrix. For instance, for the Ising
model $H(x)=-\sum_{(i,j)}x_ix_j$, $x_i=\pm 1$,
$$ \phi^e =\phi= \left(%
\begin{array}{cc}
  \sqrt{\sinh{\frac{\beta}{2}}} & \sqrt{\cosh{\frac{\beta}{2}}} \\
  -\sqrt{\sinh{\frac{\beta}{2}}} & \sqrt{\cosh{\frac{\beta}{2}}} \\
\end{array}%
\right)\;, $$ which is indeed invertible for all $\beta$.

\subsection{Injectivity and Criticality}

Surprisingly injectivity does, in the case of PEPS corresponding
to classical models, not depend on the type of interaction, but
merely on the lattice geometry of the PEPS, which coincides with
the interaction graph of the classical model. If a subset on the
interaction graph is such that every site has at most one outgoing
edge, then the corresponding region has the injectivity
property---otherwise it does not. The simple reason is as follows.

Consider a region $R$ on the interaction graph and let $E$ be the
set of edges connecting sites on the boundary $\bar{R}\subseteq R$
to exterior sites. We are interested in the injectivity of the
operator $\Gamma_R:\mathbb{C}^{D \otimes  |E|}\rightarrow
\mathbb{C}^{d\otimes |R|}$. The product form of the $A$'s
(\ref{eq:tensor-classical}) leads to
\begin{equation}\label{eq:dimension-classical} \langle
x|\Gamma_R|\bar{\alpha}\rangle = C(x)\; \langle
\bar{x}|F|\bar{\alpha}\rangle\;,\end{equation} where the bar
always means ``at the boundary'' and $C(x)$ is the result of the
contractions along all $e\not\in E$. In the case $|E|=|\bar{R}|$
we have that the vectors $\bar x$ and $\bar \alpha$ have the same
dimension and $F=\otimes_{e\in E} \phi^e$.

Hence,  $F$ is invertible so that $\Gamma_R$ is injective
\footnote{Strictly speaking this is only true if for each
$\bar{x}\in\bar{R}$ there is at least one configuration $x$ such
that $C(x)\neq 0$. This should again be true in the ``generic''
case. For the Ising model  one can indeed show analytically that
$C(x)\neq 0$ for all configurations $x$ and all values of $\beta$
(including the critical point).}. On the other hand if
$|E|>|\bar{R}|$ then $F$ is rectangular and not invertible, such
that $\Gamma_R$ cannot be injective. Hence, we do not have
injectivity for the square lattice but we do have it e.g. for the
tetrahedral 3D or the hexagonal 2D lattice (Figure
\ref{figure}.a). A 2D square lattice can as well lead to
injectivity if we either introduce a substructure at each point
(Figure \ref{figure}.c) or if we allow for defects (Figure
\ref{figure}.b): if with probability $p$ a bond is missing then
injective regions contain of the order of $(1-p)^2/p^3$ lattice
sites.

\begin{figure}[tt]
\begin{center}
\epsfig{file=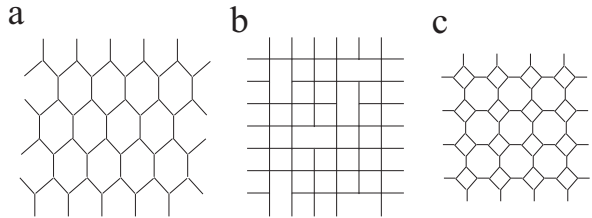,angle=0,width=0.9\linewidth}
\end{center}
\caption {Examples of 2D lattice geometries leading to injectivity
when the PEPS is constructed from a classical spin model with
corresponding interaction graph. A 3D example would be a
tetrahedral crystal lattice (ice).} \label{figure}
\end{figure}

\

With this at hand we can now give an example of an injective PEPS
whose associated parent Hamiltonian is gapless. It is the PEPS
associated to the 2D classical (isotropic) hexagonal Ising model.
On the one hand, we have injectivity and hence the PEPS is the
unique ground state of its parent Hamiltonian. On the other hand,
the Ising model becomes critical for $\beta=\frac12 \ln
(2+\sqrt{3})$ \cite{hc}, which implies that the PEPS has
correlations with power-law decay. By \cite{Hastings2}, this
implies that the parent Hamiltonian has to be gapless.

\subsection{Uniqueness for non-injective lattices}

Although injectivity is a generic and powerful condition for
proving uniqueness of the ground state it is not a necessary
requirement. In fact, in the case of a PEPS associated to a
classical model on a square lattice, although we do not have
injectivity, it is still the {\it unique} ground state of its
parent Hamiltonian \footnote{We are assuming again the generic
conditions that the matrices $\phi^e$ are invertible and that
$C(x)\not = 0$ for all configurations $x$, which is true e.g. for
the Ising model.}. The reason is that we can essentially follow
the main proof given above without using injectivity.  The only
thing we need for that is to show that the intersection properties
of the ranges $G_{R}$ in Lemma \ref{lem:crucial} remain true. This
can be shown easily just using (\ref{eq:dimension-classical}) and
some dimension considerations. We will illustrate them in a
particular case, that is, we will prove that, with the obvious
notation,
\begin{equation}\label{eq.1}
G_{\begin{diagram} \dgARROWLENGTH=0.3em
 \node{\bullet}\arrow{l,-} \arrow{s,-} \node{\bullet}\arrow{l,-} \arrow{s,-} \node{\bullet}\arrow{s,-}\\
 \node{\bullet}\arrow{l,-} \arrow{s,-}\node{\bullet}\arrow{l,-} \arrow{s,-} \node{\bullet} \arrow{s,-}\\
 \node{\bullet}\arrow{l,-}\node{\bullet}\arrow{l,-}
\node{\bullet}
\end{diagram}}\otimes \mathcal{H}\cap \mathcal{H}\otimes G_{\begin{diagram} \dgARROWLENGTH=0.3em
 \node{\bullet}\arrow{l,-} \arrow{s,-} \node{\bullet}\arrow{l,-} \arrow{s,-} \node{\bullet}\arrow{s,-}\\
 \node{\bullet}\arrow{l,-} \arrow{s,-}\node{\bullet}\arrow{l,-} \arrow{s,-} \node{\bullet} \arrow{s,-}\\
 \node{\bullet}\arrow{l,-}\node{\bullet}\arrow{l,-}
\node{\bullet}
\end{diagram}}=G_{\begin{diagram} \dgARROWLENGTH=0.3em
 \node{\bullet}\arrow{l,-} \arrow{s,-} \node{\bullet}\arrow{l,-} \arrow{s,-} \node{\bullet}\arrow{s,-}
 \arrow{l,-}\node{\bullet}\arrow{s,-}\\
 \node{\bullet}\arrow{l,-} \arrow{s,-}\node{\bullet}\arrow{l,-} \arrow{s,-}
  \node{\bullet} \arrow{s,-} \arrow{l,-}\node{\bullet}\arrow{s,-}\\
 \node{\bullet}\arrow{l,-}\node{\bullet}\arrow{l,-}
\node{\bullet}\arrow{l,-}\node{\bullet}
\end{diagram}}
,\end{equation} with each dot corresponding to a two-dimensional
Hilbert space.

Since the inclusion $\supset$ is trivial it is enough to show that
the dimension of the left hand side is $\le 2^{10}$ (that, by
(\ref{eq:dimension-classical}) is the dimension of the right hand
side). Now, the operator $\mathbbm{1}\otimes |0\>_a\<0|_a$ acting
on  $$\mathcal{H}\otimes G_{\begin{diagram} \dgARROWLENGTH=0.3em
 \node{\bullet}\arrow{l,-} \arrow{s,-} \node{\bullet}\arrow{l,-} \arrow{s,-} \node{\bullet}\arrow{s,-}\\
 \node{\bullet}\arrow{l,-} \arrow{s,-}\node{a}\arrow{l,-} \arrow{s,-} \node{\bullet} \arrow{s,-}\\
 \node{\bullet}\arrow{l,-}\node{\bullet}\arrow{l,-}
\node{\bullet}
\end{diagram}}$$
has trivial kernel, where $a$ is the position $(2,3)$. To see
this, it is enough to notice that by
(\ref{eq:dimension-classical}),
$$\dim(\mathbbm{1}\otimes |0\>_a\<0|_a(G_{\begin{diagram} \dgARROWLENGTH=0.3em
 \node{\bullet}\arrow{l,-} \arrow{s,-} \node{\bullet}\arrow{l,-} \arrow{s,-} \node{\bullet}\arrow{s,-}\\
 \node{\bullet}\arrow{l,-} \arrow{s,-}\node{a}\arrow{l,-} \arrow{s,-} \node{\bullet} \arrow{s,-}\\
 \node{\bullet}\arrow{l,-}\node{\bullet}\arrow{l,-}
\node{\bullet}
\end{diagram}}))=2^8.$$

\

In particular $\mathbbm{1}\otimes |0\>_a\<0|_a$ has also trivial
kernel when acting on the left hand side of (\ref{eq.1}).
Therefore, the dimension of this left hand side is $$\le
\dim(\mathbbm{1}\otimes |0\>_a\<0|_a(G_{\begin{diagram}
\dgARROWLENGTH=0.3em
 \node{\bullet}\arrow{l,-} \arrow{s,-} \node{\bullet}\arrow{l,-} \arrow{s,-} \node{\bullet}\arrow{s,-}\\
 \node{\bullet}\arrow{l,-} \arrow{s,-}\node{\bullet}\arrow{l,-} \arrow{s,-} \node{a} \arrow{s,-}\\
 \node{\bullet}\arrow{l,-}\node{\bullet}\arrow{l,-}
\node{\bullet}
\end{diagram}}\otimes\mathcal{H}))=2^{10},$$

\noindent where the last equality comes from the fact that, again
by (\ref{eq:dimension-classical}),
$$\dim(\mathbbm{1}\otimes |0\>_a\<0|_a(G_{\begin{diagram} \dgARROWLENGTH=0.3em
 \node{\bullet}\arrow{l,-} \arrow{s,-} \node{\bullet}\arrow{l,-} \arrow{s,-} \node{\bullet}\arrow{s,-}\\
 \node{\bullet}\arrow{l,-} \arrow{s,-}\node{\bullet}\arrow{l,-} \arrow{s,-} \node{a} \arrow{s,-}\\
 \node{\bullet}\arrow{l,-}\node{\bullet}\arrow{l,-}
\node{\bullet}
\end{diagram}}))=2^7.$$

\

Finally, it has to be noticed that, if we consider the following
regions,
$$
R_1=\; \begin{diagram}\dgARROWLENGTH=0.8em \node{\bullet}
\node{\bullet}  \node{\bullet}
 \\
\node{\bullet}  \node{\bullet} \node{\bullet}
 \\
\node{\bullet}  \node{\bullet}  \node{\bullet}
\end{diagram}\quad \quad
R_2=\;\begin{diagram}\dgARROWLENGTH=0.8em \node{}  \node{\bullet}
\node{}
 \\
\node{\bullet}  \node{\bullet} \node{\bullet}
 \\
\node{}  \node{\bullet}  \node{}
\end{diagram}\; ,
$$
by dimensional considerations using (\ref{eq:dimension-classical})
as above, it is straight forward to verify that
$P_{R_1}=P_{R_2}\otimes \1$, where $P_R$ is the projection onto
the orthogonal complement of $G_R$. Then, the parent Hamiltonian
of our PEPS can be considered to be both $P_{R_1}$ or $P_{R_2}$.
While $P_{R_1}$ has still a square structure, $P_{R_2}$ is more
natural in the sense that it reflects exactly the nearest
neighbors of the spin in the center of a cross (for this reason it
is the one that will appear in the next section).

\section{A computable sufficient condition for an energy
gap}\label{sec:gap}

The detection of either criticality or the occurrence of a
spectral gap is an important problem in both condensed matter
theory and quantum information theory. Unfortunately in the 2D
situation there are very few criteria for the existence of a gap
above the ground state energy. In this section we will provide a
computable sufficient condition for a gap in the parent
Hamiltonian of a PEPS. The idea comes from the 1D case
\cite{FaNaWe} and the proof is essentially the same. For
simplicity we consider a translational invariant frustration free
local Hamiltonian on a 2D square lattice. Let $h$ be a locally
acting projector and $h_{ij}$ its translate by $i$ columns and $j$
rows so that
 $H=\sum_{ij}h_{ij}$. Since $H$ is local, there exists a
small $I\subset \{1,\ldots,N\}\times \{1,\ldots, M\}$ such that $h
h_{ij}\ge 0$ whenever $(i,j)\not \in I$. Then,

\begin{prop}
If $H$ has a unique ground state (i.e., it is the parent
Hamiltonian of an injective PEPS) and
\begin{equation}\label{eq:gap}
\sum_{(i,j)\in I} hh_{ij} + h_{ij} h
> -\frac{1}{|I|+1}\left(h+ \sum_{(i,j)\in I} h_{ij}\right)
\end{equation}
then there is an $\epsilon>0$ such that $H^2>\epsilon H$ and hence
there is a uniform (independent of the size of the system) energy
gap for $H$.
\end{prop}

In general, one can replace the right hand side of (\ref{eq:gap})
by
\begin{equation*}
 -\frac{1}{\sum \alpha_{ij}}\left(\alpha_{00}h+ \sum_{(i,j)\in I}\alpha_{ij}
 h_{ij}\right).
\end{equation*}

In the case of the PEPS $|\psi\>$ associated to the classical
Ising model on the square lattice, the above criterion finds a gap
for $\beta<0.27$. In this case one can indeed show analytically
the existence of an energy gap for the whole range $\beta
<\beta_c=\frac{1}{2}(1+\sqrt{2})$ \cite{PEPS-Frank}.

To do so we need to use the concept of a Q-matrix: a matrix such
that $q_{ii}\le 0$ for all $i$, $q_{ij}\ge 0$ for all $i\not = j$,
and $\sum_i q_{ij} = 0$ for all $j$. As a consequence of \cite[Thm
2.1.2]{Norris} the eigenvalues of $Q$ verify $0=\lambda_0\ge
\lambda_1\ge \ldots \ge \lambda_{n-1}\ge -1$. The Q-matrix we are
going to use is the one that generates the Glauber dynamics
associated to the classical Ising model \cite{Martinelli}. That
is, for a function $f$ on the configuration space we define the
gradient of a function $f$ at the site $i$ as the function
$\bigtriangledown_if(x)=f(x^i)-f(x)$, where $x^i$ is the
configuration $x$ after flipping the spin at site $i$. With this
notation we define
$$Qf(x)=\sum_{i}c(i,x)\bigtriangledown_if(x)$$
where the sum is on the sites $i$ of the lattice and the
transition rates $c(i,x)$ are those associated to the {\it
Metropolis} dynamics
$$c(i,x)=\min\{1,e^{-\beta(\bigtriangledown_iH_i)(x)}\},$$
with $H_i(x)=-\sum_j x_ix_j$ (the sum on the $j$'s connected by
$i$ in the interaction graph; in the case of the square lattice
the nearest neighbors of $i$).

$Q$ is trivially a $Q$-matrix, and it is proven in
\cite{Martinelli} that $\gap(Q)=-\lambda_1$ has a uniform
(independent of the size of the system) lower bound. Now we
consider the Hamiltonian in our quantum system given by $H_Q=-Q$.
If we define the matrices $Q_i$ by
$$\<y|Q_i|x\>=\left\{
\begin{array}{c}
  -c(i,x)\, , \quad y=x \\
  c(i,x)\, , \quad y=x^i \\
  0\, , \quad else \\
\end{array}\right. \; ,$$
we have that
\begin{enumerate}
\item[(i)] $Q_i$ is a $Q$-matrix (and hence $\le 0$), \item[(ii)]
$Q=\sum_i Q_i$, \item[(iii)] $Q_i$ acts only on the nearest
neighbors of $i$, \item[(iv)] $Q|\psi\>=0$ \cite[Thm.
3.5.5]{Norris} and hence $Q_i|\psi\>=0$ for all $i$.
\end{enumerate}

With these properties and using (\ref{eq:tensor-classical}) it is
easy to show that $H_Q$ is in matrix ordering upper bounded by the
parent Hamiltonian of our PEPS. This immediately implies that the
parent Hamiltonian is gapped.\vspace*{5pt}

\emph{Acknowledgements: } D.P.-G. was supported by the Spanish
grants MTM2005-00082 and Ramon y Cajal.

\section{Appendix: Site-independent tensors}

Here we prove that for every PEPS which is translational invariant
on a $N\times M$ 2D square lattice there exists a PEPS
representation with site-independent tensors. The proof follows
closely the 1D case \cite{MPS-Perez}.

For doing tensor contraction we can apply Dirac notation along the
edges of the lattice, but we will need some conventions. The first
is that, when we do a horizontal (resp. vertical) contraction, the
vertical (resp. horizontal) indices always tensorize. For example,
the result from contracting the positions $(1,1)$ and $(1,2)$ will
be
$$\sum_{\scriptsize \begin{array}{c}
  l_1, u_1, d_1 \\
  r_2, u_2, d_2 \\
  a\\
\end{array}} A^{(1,1)}_{l_1, a, u_1, d_1} A^{(1,2)}_{a, r_2, u_2,
d_2 } |l_1\rangle_h\langle r_2|_h|u_1, u_2\rangle_v\langle
d_1,d_2|_v.$$ The other convention is that the edges of each row
(column) are also joined.

We start with a general PEPS:
\begin{equation*}
|\varphi\> =\sum_{i_{(1,1)},\ldots, i_{(N,M)}}
\C\left((A^{(j,k)}_{i_{(j,k)}})_{(j,k)}\right)|i_{(1,1)}\rangle
\cdots|i_{(N,M)}\rangle
\end{equation*}
and define site-independent tensors $(NM)^{\frac{1}{NM}}S_{i}$ as
$$\sum_{u,d,l,r, j, k} A^{(j,k)}_{i;u,d,l,r}|j,k,l\rangle_h\langle
j,k+1, r|_h|k,j,u\rangle_v\langle k,j+1,d|_v.$$ With this choice
translational invariance immediately implies that
$$|\varphi\>=\sum_{i_{(1,1)},\ldots, i_{(N,M)}}
\C\left((S_{i_{(j,k)}})_{(j,k)}\right)|i_{(1,1)}\rangle
\cdots|i_{(N,M)}\rangle.$$

\end{document}